\documentclass{article}

\usepackage{graphicx}
\usepackage[margin=1in]{geometry}
\usepackage{xurl}
\usepackage[colorlinks=true, urlcolor=blue, breaklinks=true]{hyperref}
\usepackage{float}
\usepackage{booktabs}
\usepackage{caption}
\usepackage{subcaption}
\usepackage{enumitem}
\usepackage{array}
\usepackage{xcolor}
\usepackage{comment}
\usepackage[super,comma]{natbib}
\usepackage{makecell}
\usepackage{amsmath}
\usepackage{tcolorbox}
\usepackage[table]{xcolor} 

\title{\textbf{Search in Transition: A Study of University Students’ Perspectives on Using LLMs and Traditional Search Engines in English Test Problem Solving for Higher Study}}

\author{
Tarek Rahman\textsuperscript{1}[0009-0006-0224-6516], 
Md Shaharia Hossen\textsuperscript{1}[0009-0009-8692-811X], \\
Mark Protik Mondol\textsuperscript{1}, 
Jannatun Noor Mukta\textsuperscript{1}[0000-0001-9669-151X] \\
\\
\textsuperscript{1}Department of Computer Science and Engineering (CSE), \\
United International University (UIU), Dhaka, Bangladesh \\
\\
trahman221182@bscse.uiu.ac.bd, \\
mhossen201288@bscse.uiu.ac.bd, mmondol221129@bscse.uiu.ac.bd, \\
 jannatun@cse.uiu.ac.bd
}

\date{}

\begin{document}

\maketitle

\section*{Abstract}
    
As Artificial Intelligence (AI) becomes increasingly integrated into education, university students preparing for English language tests are frequently shifting between traditional search engines like Google and large language models (LLMs) to assist with problem-solving. This study explores students’ perceptions of these tools, particularly in terms of usability, efficiency, and how they fit into English test preparation practices.Using a mixed-methods design, we collected survey data from 140 university students across various academic fields and conducted in-depth interviews with 20 participants. Quantitative analyses, including ANOVA and chi-square tests, were applied to assess differences in perceived efficiency, satisfaction, and overall tool preference. The qualitative results reveal that students strategically alternate between GPT and Google based on task requirements. Google is primarily used for accessing reliable, multi-source information and verifying rules, whereas GPT is favored for summarizing content, providing explanations, paraphrasing, and drafting responses for English test tasks.Since neither tool independently satisfies all aspects of English language test preparation, students expressed a clear preference for an integrated approach. In response, this study proposes a prototype chatbot embedded within a search interface, combining GPT’s interactive capabilities with Google’s credibility to enhance test preparation and reduce cognitive load.

\vspace{2\baselineskip}
\providecommand{\keywords}[1]{\par\noindent\textbf{Keywords: }#1}

\keywords{Large Language Models in Education; English Language Test Preparation; Hybrid Search and Learning Environments; Human Computer Interaction; AI-Supported Learning;Search Engine}

\section{Introduction}

The rapid advancement of artificial intelligence (AI) has substantially reshaped how university students prepare for English language tests and engage with digital language-learning resources \cite{pirzado2024navigating}. Traditionally, students have relied on search engines such as Google to access grammar rules, vocabulary explanations, and test-related materials due to their broad coverage, accessibility, and reliance on authoritative sources. More recently, the emergence of large language models (LLMs), including OpenAI’s ChatGPT, has introduced a conversational mode of interaction that provides direct explanations, summaries, and language support, potentially increasing efficiency in English language test preparation \cite{alberth2023chatgpt}.

This technological shift raises important questions about how learners perceive and utilize these tools in high-stakes English language test contexts, where accuracy, clarity, and time efficiency are critical. Prior research suggests that LLMs are particularly effective for tasks such as summarization, paraphrasing, and drafting responses, although their reliability can vary depending on task complexity and linguistic specificity \cite{divekar2024choosing,xu2023chatgpt}. In contrast, traditional search engines offer access to official test resources and rule-based explanations but often require learners to navigate multiple sources and independently evaluate the relevance and consistency of information. As a result, existing studies increasingly characterize LLMs and search engines as complementary rather than competing tools, with learners using LLMs for explanation and drafting while turning to search engines for verification and rule confirmation \cite{caramancion2024llms,spatharioti2023comparing}.

Despite the strengths of both approaches, each presents notable limitations when used in isolation. LLMs may generate fluent yet inaccurate outputs, which can mislead learners if not carefully verified \cite{xu2023chatgpt}. Conversely, traditional search engines may contribute to information overload and inefficiency, particularly under the time constraints typical of test preparation scenarios. Consequently, many students adopt a hybrid strategy that involves alternating between LLM-based tools and search engines to balance speed, fluency, and credibility \cite{sakirin2023user,kapoor2024ai}. However, frequent tool switching can increase cognitive load and disrupt learning workflows, especially during complex English language test problem-solving tasks.

To systematically examine these dynamics in the context of English language test preparation for higher studies, this study investigates students’ perceptions, usage patterns, and expectations regarding LLMs, traditional search engines, and their combined use. Specifically, the study addresses the following research questions:

\begin{itemize}
    \item \textbf{RQ1:} How do university students perceive the usability, efficiency, and satisfaction of LLM-based tools compared to traditional search engines in English language test problem solving?
    \item \textbf{RQ2:} What patterns of tool usage emerge when students perform English language test tasks using LLMs, traditional search engines, or a combination of both?
    \item \textbf{RQ3:} What are students’ preferences and expectations regarding a hybrid solution that integrates LLM-based assistance with traditional search engines for English language test preparation?
\end{itemize}

Based on a mixed-methods research design, this study makes the following contributions:

\begin{enumerate}
    \item It provides task-level empirical evidence comparing LLM-based tools and traditional search engines for English language test preparation tasks aligned with IELTS and TOEFL formats.
    \item It quantitatively demonstrates the trade-off between efficiency and accuracy across different tool usage strategies.
    \item It offers a mixed-methods analysis that captures workflow-level cognitive and usability implications of tool switching.
    \item It proposes a domain-specific, human-computer interaction (HCI)-oriented hybrid interaction model grounded in empirical user data.
\end{enumerate}

\section{Literature Review}

Large language models (LLMs) have increasingly reshaped how learners engage with language-related tasks, influencing how university students prepare for English language tests and retrieve linguistic information. While traditional search engines such as Google have long been the dominant tools for accessing grammar rules, vocabulary explanations, and test preparation materials, recent research has begun to examine how LLMs compare with search engines in terms of usability, task performance, and learner trust. Divekar et al.~\cite{divekar2024choosing} investigated how university students choose between LLMs such as ChatGPT and traditional search engines for learning purposes, finding that LLMs support rapid summarization and ease of understanding, though their effectiveness depends heavily on task complexity. Similarly, Kumar et al.~\cite{kumar2024help} demonstrated that LLM assistance can improve task formulation and learning outcomes, suggesting potential benefits for structured language tasks such as writing and paraphrasing.

Several studies have focused on task completion performance and efficiency. Spatharioti et al.~\cite{spatharioti2023comparing} conducted a randomized experiment showing that LLM users completed decision-making tasks more quickly and with fewer queries. However, they also identified a critical limitation: users often overtrust LLM outputs, particularly when incorrect responses are presented fluently and confidently. Xu et al.~\cite{xu2023chatgpt} reinforced this concern by emphasizing the importance of rigorous verification when relying on LLM-generated content. In the context of English language tests, such overreliance may negatively affect grammatical accuracy and factual correctness if learners fail to cross-check model outputs.
\begin{figure*}[t]
    \centering
    \includegraphics[width=\textwidth]{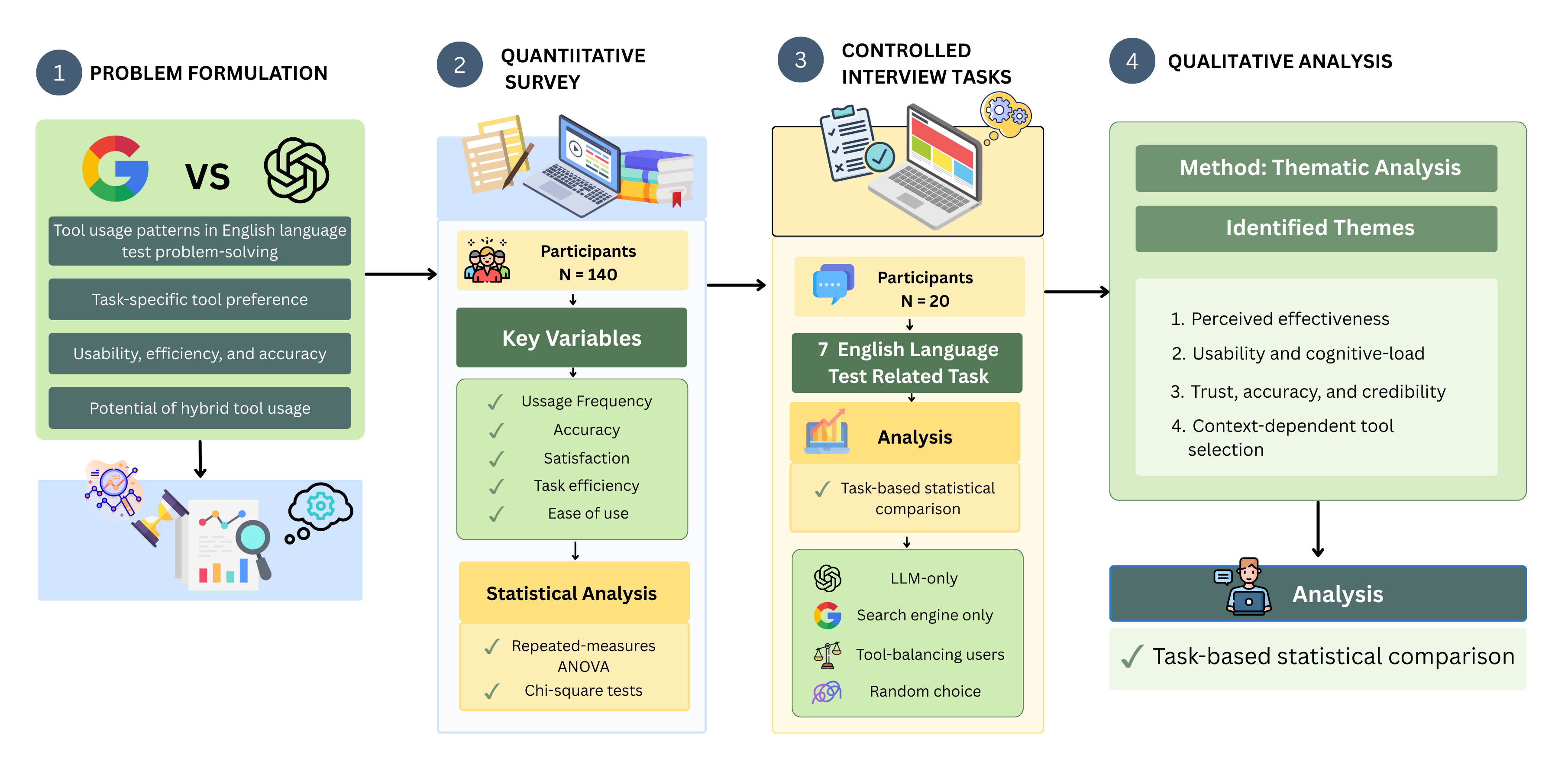}
    \caption{Overview of the study methodology. A mixed-methods approach was employed in this study. (1) The survey phase (n = 140) captured quantitative data and was analyzed using different statistical tests. (2) The qualitative phase included in-person interviews (n = 20), where participants completed seven English language test problem tasks and were grouped based on tool usage. Thematic analysis of open-ended responses and interview transcripts led to four themes.}
    \label{fig:double_column}
\end{figure*}
Research on task preference further highlights the complementary roles of LLMs and traditional search engines. Caramancion~\cite{caramancion2024llms} evaluated a range of information-seeking scenarios and found that users preferred traditional search engines for fact-based and rule-oriented queries, while favoring LLMs for complex, generative, or creative tasks. Supporting these findings, Sakirin and Said~\cite{sakirin2023user} reported that a majority of users preferred ChatGPT-style conversational interfaces due to perceived personalization, efficiency, and convenience. Extending this line of inquiry, Wazzan et al.~\cite{wazzan2024comparing} demonstrated that tool choice influences user strategy, with LLMs enabling more intuitive interactions and traditional tools requiring structured navigation patterns that are also observable in English language test preparation workflows.

Credibility and source transparency remain central concerns. Kapoor et al.~\cite{kapoor2024ai} argued that despite the speed and convenience of AI-based tools, traditional search methods remain more reliable for academic and educational contexts due to their access to authoritative sources. In contrast, LLMs often lack explicit citations, posing challenges for learners who require rule-based validation in language learning settings. To address this trade-off, researchers have proposed hybrid approaches. Early work by Bal and Nath~\cite{bal2009comparative} explored metasearch engines that aggregate multiple sources to improve accuracy, while more recent studies have advocated for systems that combine the contextual reasoning abilities of LLMs with the verification strengths of search engines~\cite{caramancion2024llms,bansal2023optimizing}.

Despite growing interest in hybrid systems, existing studies have largely examined LLMs and traditional search engines in isolation or through narrowly defined task comparisons, without fully capturing how students naturally integrate both tools during English language test preparation~\cite{xu2023chatgpt}. While tools such as Perplexity AI attempt to bridge this gap by pairing AI-generated responses with source links~\cite{perplexity2024ai}, and Google has introduced AI-powered summaries through its Search Generative Experience, these systems remain relatively static and lack personalization, real-time adaptation, and task-specific pedagogical support. Addressing these gaps, the present study employs a mixed-methods approach to examine students’ hybrid tool usage behaviors in English language test problem solving and proposes a user-informed, context-aware prototype tailored to this domain.

\section{ Methodology}
To explore university students’ preferences and usage behaviors regarding large language models (LLMs) and traditional search engines in the context of English language test problem solving for higher studies, we employed a mixed-methods approach that combined quantitative and qualitative data collection and analysis, as illustrated in Figure 1. This design enabled the examination of both broad usage patterns and deeper user experiences in a complementary manner.

We first conducted an online survey that collected responses from 140 university students across a range of academic disciplines. While the survey primarily targeted students from technology-related fields such as Computer Science and Engineering (CSE), Electrical and Electronics Engineering (EEE), and Data Science, it was also distributed to students from other disciplines, including Business Administration and English studies, to ensure diversity. The questionnaire included both closed-ended and open-ended questions designed to assess tool usage frequency, satisfaction, efficiency, and perceived ease of use when employing GPT-based LLMs and traditional search engines such as Google for English language test preparation. Descriptive statistics were used to summarize the data, and inferential statistical analyses, including one-way ANOVA and chi-square tests, were conducted to evaluate differences in user perceptions and the influence of demographic variables on tool preference.

To enrich and validate the survey findings, we conducted in-depth, in-person interviews with 20 university students from CSE, EEE, BBA, and English backgrounds, primarily recruited from United International University (UIU), Dhaka, Bangladesh. Each participant was asked to complete seven English language test–related tasks representative of higher studies preparation. These tasks included summarizing an academic passage, correcting grammatical errors, selecting appropriate vocabulary in context, paraphrasing sentences, drafting a short opinion-based response similar to IELTS Writing Task~2, drafting a response similar to IELTS Writing Task 1, and interpreting information from a short academic text.

Participants were categorized into four groups based on their tool usage behavior during task completion: LLM-only users, search engine–only users, balanced users who alternated between both tools, and free-choice users who switched between LLMs and search engines based on task demands.

The qualitative analysis integrated insights from both open-ended survey responses and interview transcripts. A thematic coding approach was applied, focusing on perceived effectiveness, usability, trustworthiness, and contextual factors influencing tool selection. This analysis offered a comprehensive understanding of how students navigate the strengths and limitations of LLMs and traditional search engines in English language test problem solving, and how their choices are shaped by task type, familiarity with English language rules, and perceived cognitive effort.
\section{ Demographics}
\subsection{Survey}

The survey included \textbf{140 university students} from a range of academic disciplines and demographic backgrounds. While the majority of participants were from Computer Science and Engineering (CSE), the sample also included students from Electrical and Electronics Engineering (EEE), Business Administration (BBA), Data Science, English, and other disciplines. Table~\ref{tab:demographics} summarizes the distribution of participants by department, CGPA range, gender, and age group.

The survey instrument consisted of Likert-scale questions designed to assess students’ perceptions and usage of traditional search engines (e.g., Google) and LLM-based tools (e.g., ChatGPT) in the context of \textbf{English language test preparation for higher studies}. Participants responded using a five-point scale: \emph{Never}, \emph{Rarely}, \emph{Occasionally}, \emph{Frequently}, and \emph{Always}. Each set of questions was presented in parallel for both tool categories and covered four core dimensions:

\begin{itemize}
    \item Frequency of tool usage for English language test tasks
    \item Satisfaction with the accuracy of information provided by the tools
    \item Perceived efficiency in completing English language test problems
    \item Ease of use for English language test preparation
\end{itemize}

Participants rated each item separately for traditional search engines and LLM-based tools. At the end of the survey, respondents were also asked to indicate their overall preference among search engines, LLMs, or a hybrid use of both tools.

This combination of parallel metrics and comparative judgment enabled consistent statistical comparisons between tools, while the final preference item provided insight into students’ overall inclinations toward tool usage in English language test problem solving.

\begin{table}[t]
\centering
\begin{tabular}{lr}
\hline
\multicolumn{2}{c}{\textbf{Department}} \\
\hline
Computer Science \& Engineering (CSE) & 72 \\
Electrical \& Electronics Engineering (EEE) & 24 \\
Business Administration (BBA) & 18 \\
Data Science & 12 \\
English & 10 \\
Other & 4 \\
\hline
\multicolumn{2}{c}{\textbf{CGPA Range}} \\
\hline
3.81--4.00 & 25 \\
3.51--3.80 & 29 \\
3.01--3.50 & 62 \\
2.50--3.00 & 24 \\
\hline
\multicolumn{2}{c}{\textbf{Gender}} \\
\hline
Male & 93 \\
Female & 47 \\
\hline
\multicolumn{2}{c}{\textbf{Age Range}} \\
\hline
18--20 years & 9 \\
21--25 years & 128 \\
26--30 years & 3 \\
\hline
\end{tabular}
\caption{Distribution of participants by department, CGPA range, gender, and age group}
\label{tab:demographics}
\end{table}

\subsection{In-person Interview}

To complement the survey findings and gain deeper insights into students’ tool usage behaviors, we conducted in-person interviews with \textbf{20 university students} from diverse academic backgrounds, including Computer Science and Engineering (CSE), Electrical and Electronics Engineering (EEE), Business Administration (BBA), and English studies backgrounds, primarily recruited from United International University (UIU), Dhaka, Bangladesh.

The interview protocol consisted of a structured sequence of \textbf{seven English language test–related tasks} designed to reflect common activities involved in preparation for higher studies. Participants completed seven English language test–related tasks:
(1) summarizing an english languages test based,
(2) correcting grammatical errors in a language testing problem,
(3) selecting appropriate vocabulary in context,
(4) paraphrasing sentences without altering meaning,
(5) drafting an opinion-based response similar to IELTS Writing Task 2,
(6) drafting a data description response similar to IELTS Writing Task 1, and
(7) interpreting information from an academic reading passage.
 The tasks were selected through consultation with English language instructors and a review of common IELTS and TOEFL preparation materials to ensure contextual relevance and varying cognitive demands. The primary objective was to observe how tool choice influenced task strategy, accuracy, and efficiency across different English language test activities.

To assess task performance, we developed task-specific evaluation rubrics in consultation with faculty members specializing in English language teaching. For instance, summarization tasks were evaluated based on coherence, coverage, and conciseness; grammar and vocabulary tasks were assessed based on correctness and appropriateness; and writing tasks were scored according to clarity, coherence, and task fulfillment. Each task was independently evaluated by two raters to ensure inter-rater reliability.

Participants were categorized into four groups based on their tool usage behavior during task completion: (1) \textbf{LLM-only users}, (2) \textbf{search engine–only users}, (3) \textbf{tool-balancing users} who sequentially used both LLMs and search engines, and (4) \textbf{free-choice users} who selected tools flexibly depending on task demands. This grouping enabled comparisons of accuracy and task completion time across different task types and facilitated analysis of how tool-switching behavior related to user preferences and task complexity.

In addition to task performance, the interviews included open-ended reflections on perceived usability, trustworthiness, and strengths or limitations of each tool. These qualitative responses were analyzed using thematic coding and were used to contextualize quantitative findings and inform the design considerations for a hybrid English language test support system.

\section{Quantitative Analysis}
\subsection{ Survey Results}
Closed-ended survey responses were converted into numerical values for analysis using a five-point Likert scale coded as follows: \emph{Never} (0), \emph{Rarely} (1), \emph{Occasionally} (2), \emph{Frequently} (3), and \emph{Always} (4). This encoding enabled the computation of descriptive statistics, including means, medians, modes, and standard deviations, across four core dimensions: usage frequency, satisfaction, efficiency, and ease of use, for both traditional search engines and LLM-based tools.

\begin{figure}[t]
    \centering
    \includegraphics[width=\columnwidth]{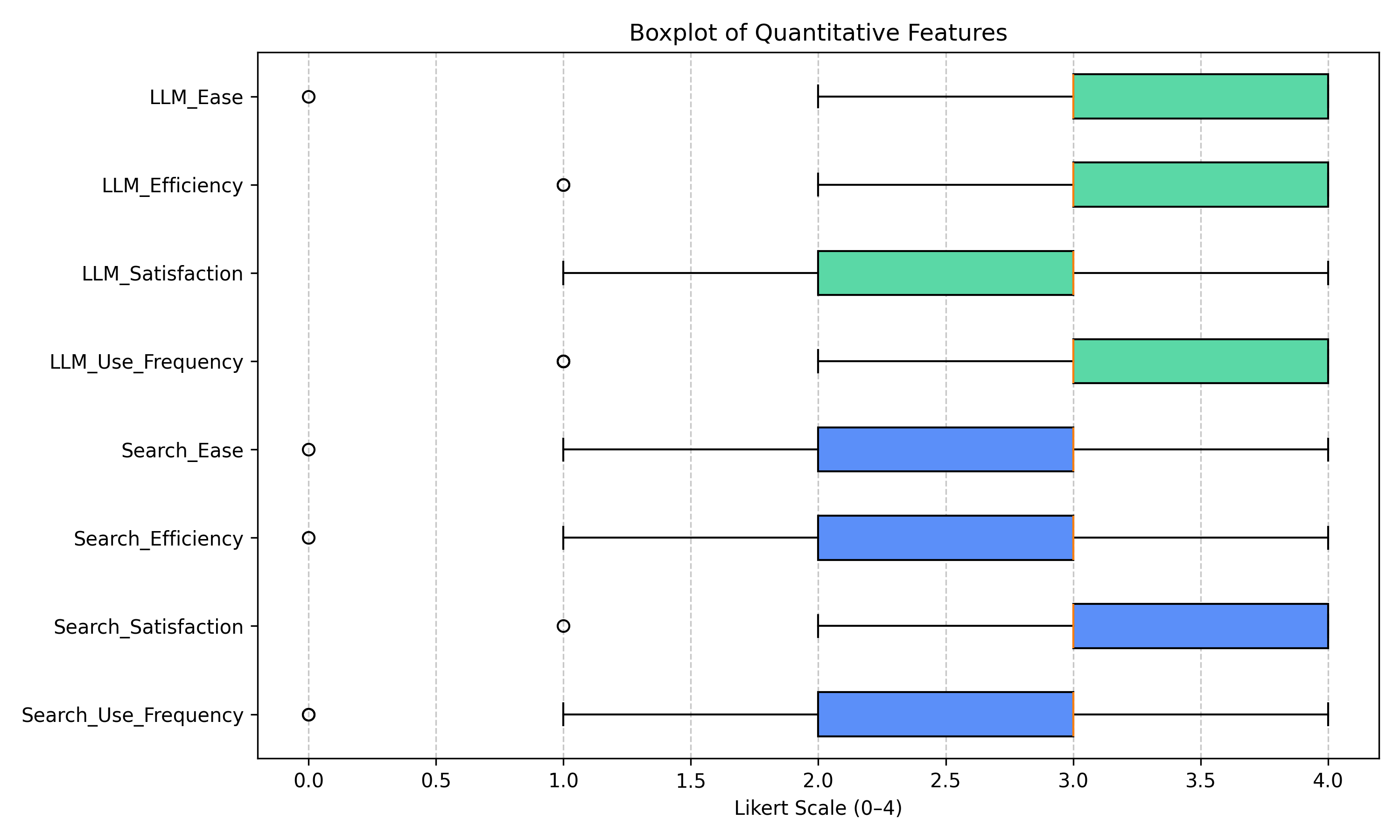}
        \caption{Boxplot of quantitative features. This figure presents a comparative analysis of key usability dimensions, usage frequency, satisfaction, efficiency, and ease of use between traditional search engines and LLM-based tools in the context of English language test problem-solving for higher studies. The top four features (Search\_Use\_Frequency, Search\_Satisfaction, Search\_Efficiency, and Search\_Ease) represent participant responses related to traditional search engines, while the bottom four features (LLM\_Use\_Frequency, LLM\_Satisfaction, LLM\_Efficiency, and LLM\_Ease) correspond to user experiences with large language models.}
    \label{fig:boxplot}
\end{figure}

Overall, the descriptive analysis indicated that LLM-based tools were used more frequently and received more favorable ratings across all evaluated dimensions. Traditional search engines exhibited lower mean scores across usage frequency, satisfaction, efficiency, and ease of use, whereas LLM-based tools consistently showed higher central tendency values. These trends are visually illustrated in Figure \ref{fig:boxplot}, which presents boxplots comparing user responses for search engines and LLMs across the four dimensions.

To determine whether the observed differences between traditional search engines and LLM-based tools were statistically significant, a series of one-way repeated-measures ANOVA tests was conducted across the four dimensions. This within-subjects design was appropriate, as each of the \textbf{140 participants} evaluated both tool categories, enabling direct paired comparisons.
The results revealed statistically significant differences in all examined dimensions. Specifically, usage frequency differed significantly between the two tools, $F(1, 139) = 17.05$, $p < 0.001$. A significant difference was also observed for satisfaction, $F(1, 139) = 18.63$, $p < 0.001$. Efficiency ratings showed a similarly significant effect, $F(1, 139) = 8.87$, $p = 0.003$, indicating that participants perceived LLM-based tools as more efficient than traditional search engines. Finally, ease of use demonstrated a significant difference, $F(1, 139) = 17.04$, $p < 0.001$.

Across all four dimensions, participants consistently rated LLM-based tools higher than traditional search engines. The direction and magnitude of the F-values indicate a robust and systematic preference for LLM-based tools in English language test problem solving. All statistical tests were conducted independently for each variable, and assumptions of normality and homogeneity of variance were examined and found to be satisfactory.

To examine whether overall tool preference was influenced by participant demographics, a chi-square test of independence was conducted between tool preference (LLM-based tools, search engines, or hybrid use) and demographic variables, including age group, gender, and academic department. The analysis yielded a non-significant result, $\chi^2(7, N = 140) = 2.01$, $p = 0.91$, indicating no statistically significant association between demographic factors and tool preference.

In summary, the survey findings demonstrate a clear and consistent preference for LLM-based tools over traditional search engines across all major usability dimensions. While demographic characteristics did not significantly influence tool preference, both descriptive and inferential analyses confirm that LLM-based tools are perceived as more usable, efficient, and satisfying for English language test problem solving.

\subsection{ In-person Interview}

To analyze the data collected from the in-person interviews, we examined both quantitative and qualitative aspects of participant performance while completing a series of structured English language test–related tasks. A total of \textbf{20 students} participated in this phase of the study. Each participant completed seven tasks representative of common English language test preparation activities, designed to capture a range of cognitive demands, including comprehension, grammatical accuracy, paraphrasing, and written expression.

Participants were grouped according to their tool usage strategy during task completion: \textbf{LLM-only users}, \textbf{search engine–only users}, \textbf{tool-balancing users} who used both tools sequentially, and \textbf{random-choice users} who selected tools flexibly depending on task requirements. Task performance was evaluated using a predefined rubric, and completion time was recorded for each participant. Table~\ref{tab:interview_performance} summarizes the key quantitative findings from the in-person interview study.
\begin{table}[t]
\centering
\begin{tabular}{lcc}
\hline
\textbf{Tool Usage Group} & \textbf{Accuracy (\%)} & \textbf{Time (min)} \\
\hline
LLM-only & 88 & 24 \\
Search engine-only & 80 & 31 \\
Tool-balancing & 90 & 35\\
Random-choice & 82--88 & 25--32 \\
\hline
\end{tabular}
\caption{Summary of task performance by tool usage group}
\label{tab:interview_performance}
\end{table}

\subsubsection{ Accuracy Analysis}

Each participant’s response was manually evaluated using a predefined scoring rubric tailored to the specific English language test task type. For example, summarization tasks were assessed based on coverage, conciseness, and coherence; grammar and vocabulary tasks were evaluated for correctness and appropriateness; and writing tasks were scored according to clarity, organization, and task fulfillment. This rubric-based evaluation ensured consistency and objectivity across all assessments.

Participants who relied exclusively on LLM-based tools
achieved an average accuracy of \textbf{88\%}, indicating that LLMs were effective in generating structured and fluent responses, particularly for tasks involving summarization, paraphrasing, and short-form writing. In comparison, participants who relied solely on traditional search engines attained a lower average accuracy of \textbf{80\%}. This performance gap can be attributed to the additional effort required to locate relevant information, synthesize content from multiple sources, and reformulate responses independently.

The highest performance was observed among participants who adopted a \textbf{tool-balancing strategy}, combining LLMs and search engines in a complementary manner. This group achieved an average accuracy of \textbf{90\%}, benefiting from the use of LLMs for content generation and explanation, alongside search engines for verification and rule confirmation. 

Participants in the random-choice group, who selected tools flexibly based on task demands, achieved accuracy scores ranging from \textbf{82\% to 88\%}. Variations within this group were influenced by task complexity and the appropriateness of tool selection, with higher accuracy observed when tool choice aligned effectively with the task requirements.

Overall, these findings indicate that while both LLMs and traditional search engines can support English language test problem solving, a balanced and strategic combination of the two tools yields the highest accuracy across diverse task types.
\begin{figure}[t]
    \centering
    \includegraphics[width=\columnwidth]{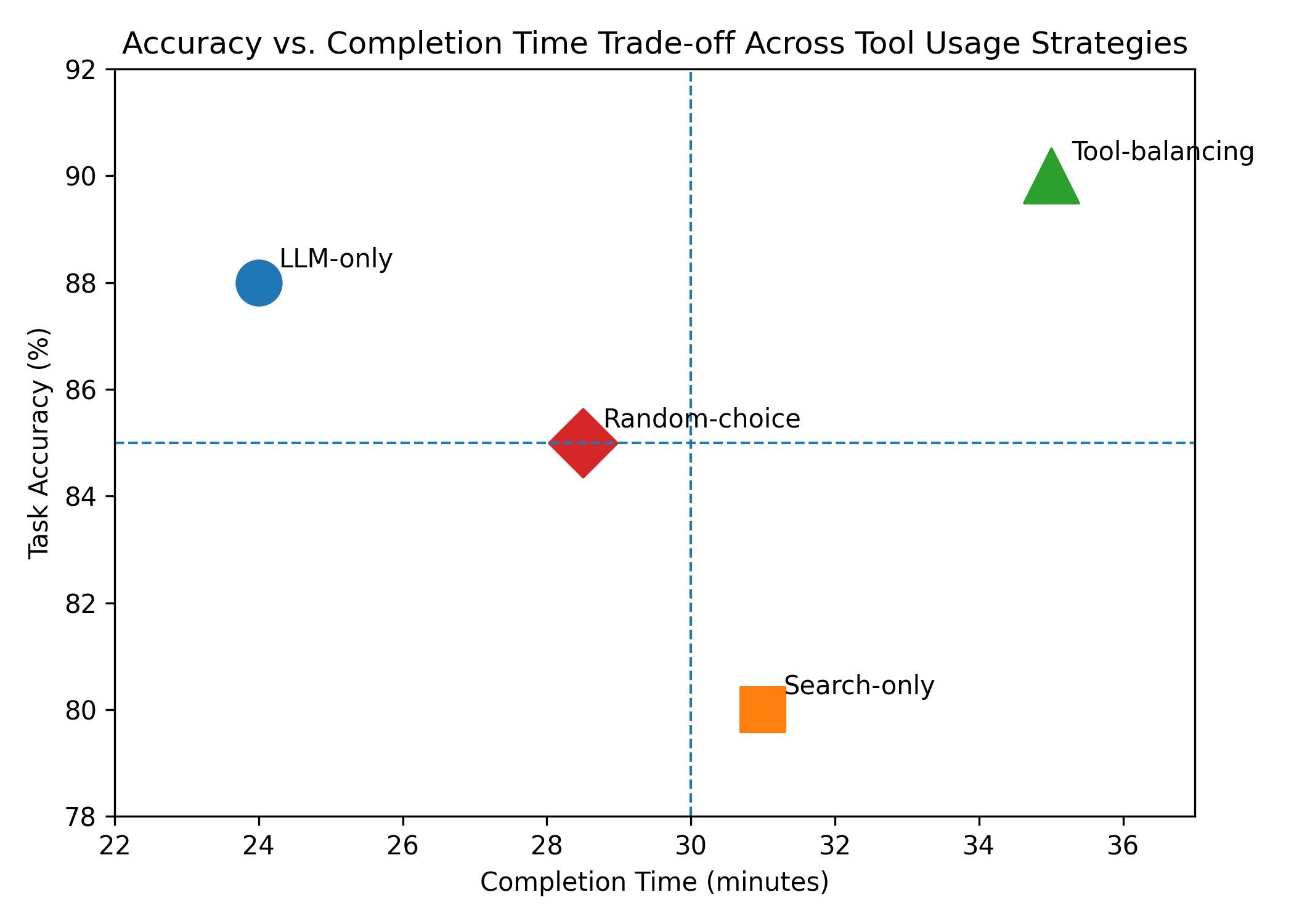}
    \caption{Trade-off between task accuracy and completion time across different tool usage strategies in English language test problem solving.}
    \label{fig:acc_time}
\end{figure}

\subsubsection{Completion Time Analysis}

In addition to accuracy, we analyzed the time required by participants to complete the assigned English language test–related tasks. Completion times varied considerably across tool usage strategies, reflecting differences in interaction style and task approach. 

Participants who relied exclusively on LLM-based tools completed the tasks relatively quickly, with an average completion time of approximately \textbf{24 minutes}. This efficiency can be attributed to the conversational and integrative nature of LLMs, which allows users to obtain explanations, examples, and draft responses without navigating multiple external sources. 

In contrast, participants who depended solely on traditional search engines required more time, averaging around \textbf{31 minutes}. This increase in completion time is likely due to the iterative process of searching, evaluating multiple webpages, and synthesizing information manually before producing a final response.

Participants who adopted a \textbf{tool-balancing strategy}, using both LLMs and search engines in combination, exhibited the longest average completion time of approximately \textbf{35 minutes}. Although this approach was more time-consuming, it enabled participants to cross-check information and refine their responses, which corresponded with the highest accuracy scores observed in the interview study.

The \textbf{random-choice group} demonstrated the greatest variability in completion time, ranging from \textbf{25 to 32 minutes}. This variation appeared to be influenced by task complexity as well as individual familiarity with the selected tools. 

Overall, the findings suggest a clear trade-off between efficiency and performance. While LLM-based tools offer faster task completion, integrating them with traditional search engines can enhance accuracy and reliability in English language test problem solving, albeit at the cost of increased task duration.

Figure \ref{fig:acc_time} further illustrates the trade-off between task accuracy and
completion time across different tool usage strategies. While LLM-only
Users completed tasks more quickly, and tool-balancing users achieved the
highest accuracy at the cost of increased completion time.

\subsubsection{Inter Rater Reliability}
To ensure consistency in task evaluation, two independent raters assessed participant responses using predefined rubrics. Inter-rater reliability was measured using Cohen’s kappa, yielding $\kappa = 0.82$, which indicates strong agreement between evaluators. Discrepancies were resolved through discussion to reach a consensus.

\subsection{Qualitative Analysis}

The qualitative analysis draws on participants’ open-ended survey responses and in-person interview transcripts to examine how university students perceive, experience, and strategically select between LLM-based tools and traditional search engines when preparing for English language tests for higher studies. A thematic analysis approach was employed to identify recurring patterns across the qualitative data. Two researchers independently conducted initial line-by-line coding of all textual responses. These codes were subsequently discussed, compared, and iteratively refined into higher-level themes until consensus was achieved, ensuring analytical rigor and reliability.

Four dominant themes emerged from the analysis: (1) task suitability and tool selection, (2) perceptions of reliability and accuracy, (3) workflow efficiency and cognitive load, and (4) usability and interaction experience.

\paragraph{Task Suitability and Tool Selection.}
Participants consistently differentiated between LLMs and search engines based on the nature of the English language test task. LLM-based tools were frequently described as particularly useful for tasks such as summarization, paraphrasing, grammar correction, and drafting short written responses. Participants noted that LLMs were especially helpful under time constraints, as they provided immediate and structured outputs that could be refined further. In contrast, traditional search engines were preferred for tasks requiring detailed rule verification, exposure to multiple examples, or consultation of authoritative sources. Several participants emphasized that search engines were more suitable when confirming grammar rules or reviewing multiple perspectives before finalizing an answer.

\paragraph{Perceptions of Reliability and Accuracy.}
Trust emerged as a central factor influencing tool choice. While participants appreciated the fluency and coherence of LLM-generated responses, many expressed reservations regarding their factual reliability, particularly for nuanced grammar rules or language-specific explanations. Several respondents reported instances in which LLM outputs appeared convincing but required additional verification. Consequently, traditional search engines were frequently used as a secondary validation mechanism. Participants generally viewed search engines as more dependable for fact-checking and accessing credible sources, although some noted difficulties in evaluating source quality or resolving contradictory information.

\paragraph{Workflow Efficiency and Cognitive Load.}
LLM-based tools were widely perceived as effective in reducing cognitive effort, particularly under time pressure. Participants highlighted that LLMs helped them avoid navigating multiple web pages by providing concise explanations or draft responses in a single interaction. However, students who combined both tools acknowledged that switching between LLMs and search engines increased overall task duration. Despite this, they reported that the additional effort often led to improved understanding and higher-quality outputs. This hybrid workflow was especially common for complex English language test tasks involving structured writing, reading comprehension, or vocabulary usage.

\paragraph{Usability and Interaction Experience.}
Participants frequently described LLMs as interactive and supportive, likening them to a conversational tutor that guided them through problem-solving steps. This conversational interaction was perceived as particularly beneficial for clarifying misunderstandings and exploring alternative phrasing. In contrast, traditional search engines were characterized as familiar and stable but less interactive, requiring greater user initiative to locate and synthesize relevant information. Interface familiarity, preferred learning style, and comfort with open-ended search influenced tool preference, especially among students with lower confidence in English language proficiency.

Overall, participants did not view LLM-based tools and traditional search engines as direct substitutes but rather as complementary components within their English language test preparation workflow. LLMs were valued for their speed, language generation capabilities, and explanatory support, while search engines remained essential for verification, rule confirmation, and access to credible sources. Tool selection was highly context-dependent, shaped by task type, prior knowledge, and the user’s need for either efficiency or accuracy. These findings highlight the nuanced strategies students adopt when navigating digital information tools and underscore the importance of integrated systems that balance convenience with reliability.
\subsubsection{Reliability Analysis}

The internal consistency of the survey instrument was assessed using Cronbach’s alpha. The overall reliability of the scale was $\alpha = 0.81$, indicating good internal consistency. Subscale reliability values were also satisfactory, with usage frequency ($\alpha = 0.78$), satisfaction ($\alpha = 0.83$), efficiency ($\alpha = 0.74$), and ease of use ($\alpha = 0.86$). These results confirm the reliability of the survey measures used in this study.

\section{Discussion}

The combined findings from the survey and in-person interviews reveal a complex and context-sensitive relationship between LLM-based tools and traditional search engines in English language test problem-solving for higher studies. Rather than exhibiting a clear preference for a single tool, students demonstrated adaptive strategies that varied according to task requirements, time constraints, and confidence in the information needed.

LLM-based tools were widely appreciated for their ability to generate structured, coherent, and contextually appropriate responses. Participants frequently highlighted the usefulness of conversational interfaces for tasks such as summarization, paraphrasing, grammar correction, and short-form writing, where rapid feedback and reduced cognitive effort were particularly beneficial. These strengths were reflected in higher usability ratings and faster task completion times for LLM-only users. However, despite these advantages, participants remained cautious about relying solely on LLMs due to concerns regarding occasional inaccuracies, generalized explanations, or the absence of explicit source attribution. As a result, many students reported verifying LLM-generated outputs through external sources.

Traditional search engines, most notably Google, continued to play a critical role in tasks that demanded depth, credibility, and source verification. Participants emphasized the importance of accessing authoritative websites, academic resources, and multiple viewpoints, particularly for grammar rules, reading comprehension, and fact-based validation. At the same time, students reported challenges associated with traditional search engines, including information overload, fragmented content, and the cognitive effort required to compare and synthesize information across multiple links. These limitations often led to longer task completion times and increased mental effort.

\begin{figure}[t]
    \centering
    \includegraphics[width=\columnwidth]{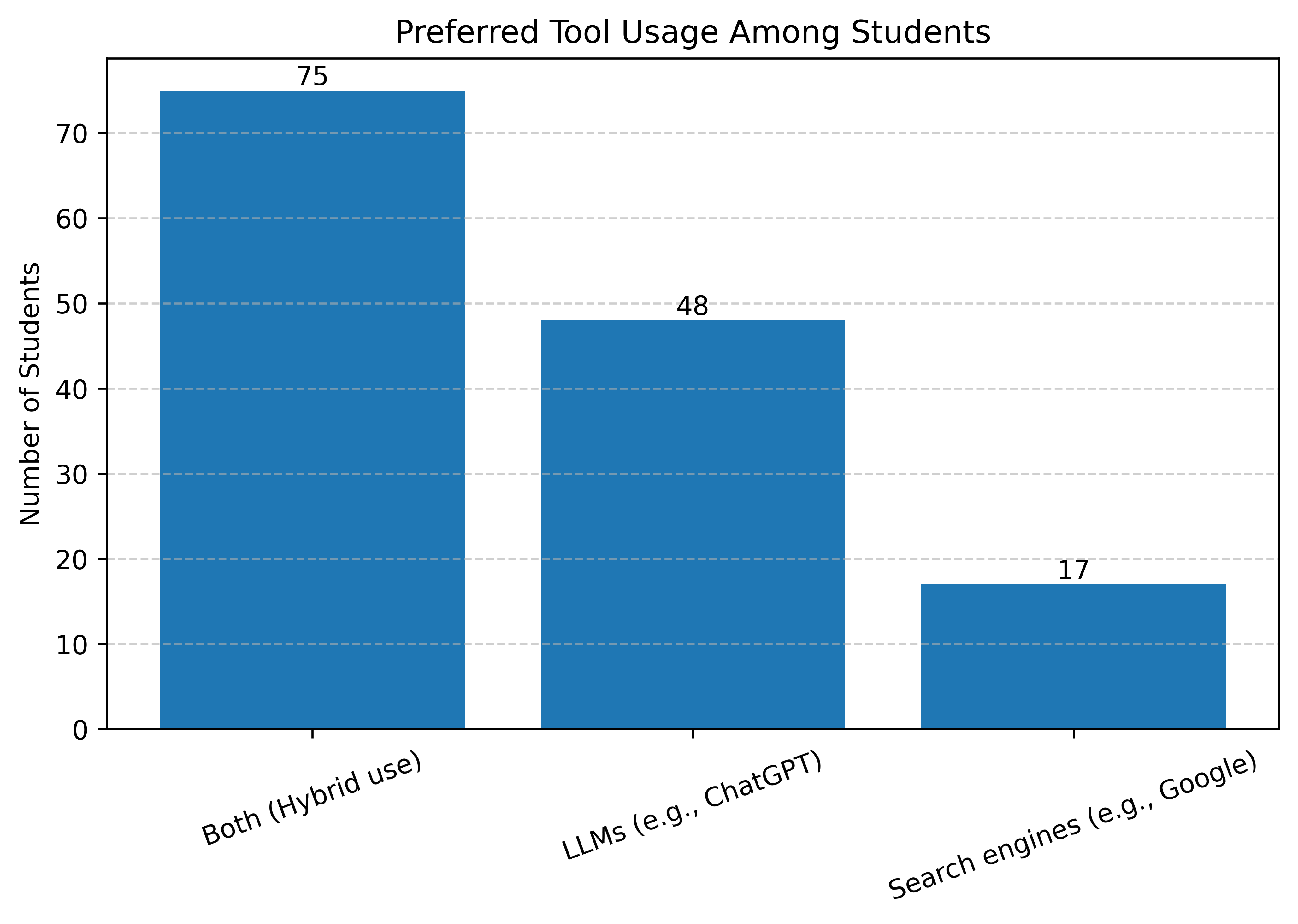}
    \caption{Distribution of participants’ overall tool preference for English language test problem solving. The majority of students preferred a hybrid approach combining LLM-based tools and traditional search engines, followed by exclusive use of LLMs, while comparatively few relied solely on traditional search engines.}
    \label{fig:preferred_tool}
\end{figure}

A central insight emerging from both quantitative and qualitative data was students’ strong inclination toward a hybrid approach that integrates the strengths of both LLMs and traditional search engines. As illustrated in Figure \ref{fig:preferred_tool}, the majority of participants expressed a preference for using both tools in combination rather than relying exclusively on either one. This preference reflects a desire for systems that support efficient content generation while maintaining access to credible, verifiable information sources.

Evidence from the interview study further reinforces the value of this integrated strategy. Participants who adopted a tool-balancing approach using LLMs for rapid explanation and drafting and search engines for verification and refinement achieved the highest task accuracy (90\%). Although this approach required more time than using a single tool, participants consistently viewed the additional effort as a worthwhile trade-off. They reported greater confidence in their final responses, improved understanding of the material, and higher overall output quality. In contrast, participants who relied exclusively on either LLMs or search engines demonstrated lower accuracy, highlighting the limitations of single-tool reliance.

Taken together, these findings suggest that students do not perceive LLMs and traditional search engines as competing technologies but rather as complementary components of an effective academic workflow. LLMs excel in synthesis, explanation, and language generation, while search engines provide depth, reliability, and access to authoritative sources. The strategic integration of these capabilities represents a promising direction for the design of future academic support systems.

More broadly, the results underscore the importance of developing intelligent, context-aware hybrid tools that reduce cognitive load without compromising academic rigor. By minimizing unnecessary tool-switching and embedding verification mechanisms within generative interfaces, such systems have the potential to enhance learning efficiency, accuracy, and trust in AI-assisted academic problem solving.
\begin{figure}[t]
    \centering
    \includegraphics[width=\columnwidth]{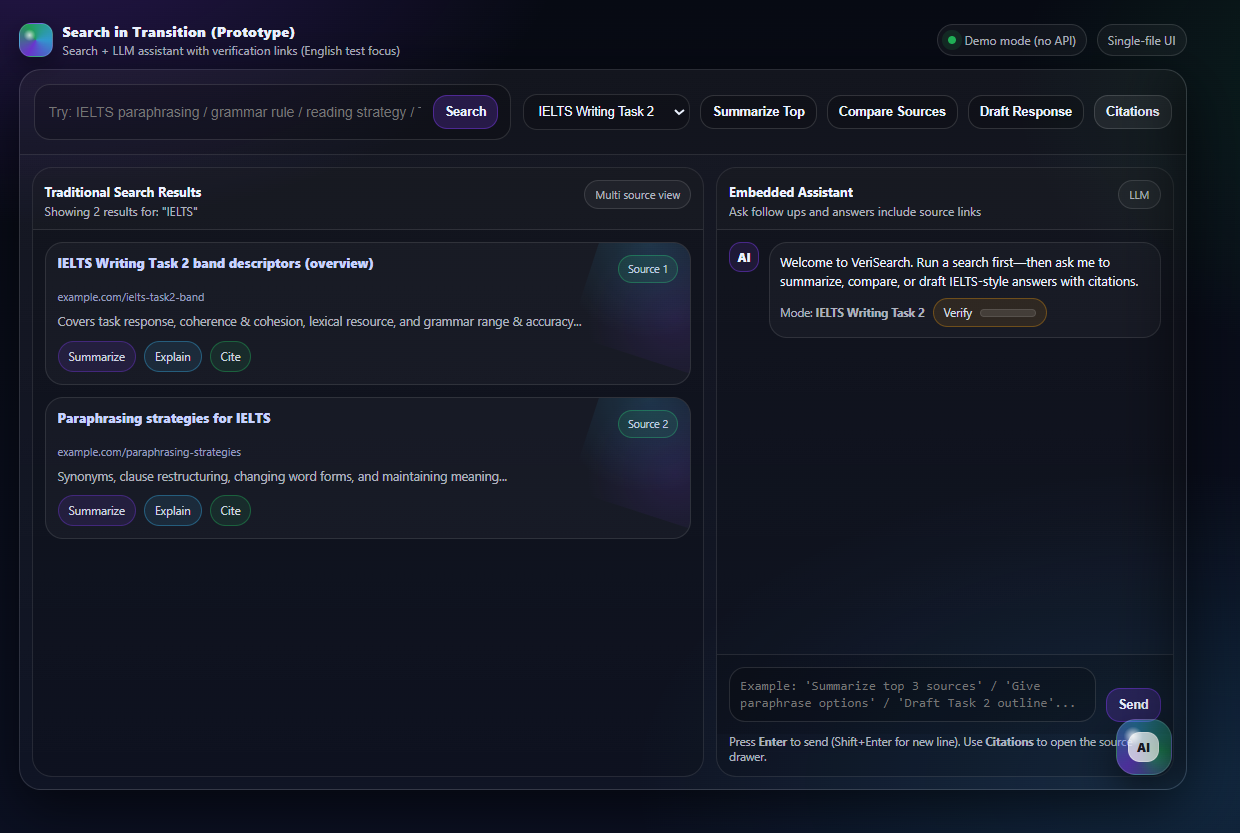}
    \caption{Conceptual prototype of the proposed hybrid search–LLM interface
for English language test preparation. The interface integrates traditional
search results with an embedded LLM assistant that provides explanations,
summaries, and draft responses, while explicitly linking generated outputs
to underlying sources to support verification and reduce cognitive load.}
    \label{fig:Conceptual_prototype}
\end{figure}
\subsection{Proposed Prototype}

Drawing on user insights from both the survey responses and the in-person interviews, we propose a conceptual, HCI-oriented hybrid prototype that embeds LLM-based assistance directly within a traditional search engine interface. The primary objective of this design is to reduce cognitive effort and workflow fragmentation caused by frequent tool switching by offering a unified interaction environment that combines the conversational utility of LLMs with the source-rich infrastructure of search engines. The design specifically targets English language test problem-solving for higher studies, reflecting the dominant use cases identified in this study.

The prototype is envisioned as an unobtrusive embedded chatbot positioned within the search interface, allowing users to engage in context-aware dialogue without disrupting their familiar browsing behavior. Unlike standalone LLM interfaces, the assistant does not replace standard search results. Instead, it complements them by offering concise summaries, follow-up clarifications, and cross-source synthesis derived from retrieved content. For example, after a query related to English language test preparation, the assistant can summarize the top search results, compare key points across multiple sources, and simplify explanations for grammar rules, reading passages, or writing tasks. Users may then ask follow-up questions to refine or extend the information without repeatedly navigating and interpreting multiple webpages.

A key contribution of the design lies in its hybrid interaction model, which enables users to move seamlessly between raw search content and AI-supported interpretation. To address concerns regarding LLM trustworthiness raised by participants, each AI-generated response is explicitly accompanied by links to the underlying sources. This design choice promotes transparency, supports verification, and mitigates the risk of unverified or hallucinated outputs in academic contexts.

Although conceptual in its current form, the prototype is grounded in the empirical findings of this study and reflects users’ demonstrated preference for combining LLM-based tools and traditional search engines. The prototype is intended to illustrate interaction design concepts rather than serve as a fully implemented system. By embedding an assistive layer into the search experience, the design aims to reduce cognitive load, increase task efficiency, and encourage evidence-based academic practices by bridging the gap between content generation and source verification. Figure~\ref{fig:Conceptual_prototype} shows the conceptual prototype, and a demonstration version of the conceptual prototype interface is publicly \href{https://search-in-transition.netlify.app/}{available}. to support transparency and reproducibility.

\section{Conclusion}
This study examined how university students combine LLM-based tools and traditional search engines when engaging in English language test problem-solving for higher studies. The findings demonstrate that students do not rely on a single tool but instead adopt complementary strategies, using LLMs for rapid explanation, paraphrasing, and drafting, while depending on traditional search engines for verification, rule confirmation, and access to credible sources. Both the survey and interview results highlight the task-dependent nature of tool selection and emphasize that efficiency, accuracy, and trust collectively shape user preferences.

The mixed-methods approach revealed a strong preference for hybrid usage patterns, with participants who strategically combined both tools achieving the highest task accuracy, despite longer completion times. These results suggest a clear trade-off between speed and reliability and underscore the limitations of relying exclusively on either LLMs or search engines in isolation. Students’ expressed concerns regarding trust and verification further reinforce the need for systems that balance generative assistance with source transparency.

Guided by these empirical insights, this study proposed a conceptual, HCI-oriented hybrid prototype that integrates LLM-based assistance within a traditional search interface. Although the prototype is conceptual and the interview sample size is limited, the findings offer practical design implications for developing AI-assisted academic tools that reduce cognitive load while supporting evidence-based learning practices.

Future work will focus on expanding participant diversity, further validating the identified qualitative themes, and implementing a functional version of the proposed prototype. Such efforts will enable controlled evaluations of usability, learning outcomes, and user trust in real-world academic settings, contributing to the development of responsible and human-centered AI systems for educational support.

\section{Limitations}

While this study offers valuable insights into university students’ use of LLM-based tools and traditional search engines for English language test problem solving, several limitations should be acknowledged. First, although the survey included a relatively large number of participants (n = 140), the in-person interview phase involved a smaller sample (n = 20). In addition, interview participants were primarily drawn from technology-related and business disciplines within a single university context, which may limit the generalizability of the qualitative findings to students from other academic backgrounds or institutional settings.

There is also potential sampling bias, as a substantial proportion of participants were enrolled in Computer Science and Engineering (CSE) programs. Consequently, the reported tool usage patterns and preferences may not fully represent students with different levels of technical familiarity or academic demands. Furthermore, the survey relied on self-reported measures of usage, satisfaction, and efficiency, which may be subject to recall bias or social desirability effects.

Although the qualitative analysis identified meaningful and consistent themes, formal inter-coder reliability metrics were not computed. The thematic agreement was achieved through discussion and consensus, which, while common in exploratory qualitative studies, may affect the replicability of the findings. Finally, the proposed hybrid prototype remains conceptual and has not yet been implemented or empirically evaluated in real academic settings. As a result, its practical effectiveness, usability, and impact on learning outcomes remain to be validated.

The study did not explicitly control for participants’ prior experience with AI tools or variations in digital literacy, which may influence tool selection and performance. Future work should incorporate these factors into controlled experimental designs.

\section{Ethical Considerations}

All research activities involving human participants in this study adhered to established ethical guidelines for academic research. Participation in both the online survey and the in-person interview sessions was entirely voluntary. Prior to data collection, all participants were informed about the purpose of the study and provided informed consent.

To protect participant privacy, no personally identifiable information was collected, stored, or analyzed at any stage of the research process. All responses were anonymized, and the data were analyzed only in aggregated form to ensure confidentiality and prevent individual identification. Participants were also informed that they could withdraw from the study at any time without penalty.

The study involved non-sensitive topics related to learning behaviors and technology usage and did not include vulnerable populations, medical procedures, or interventions. As such, the level of risk posed to participants was minimal.

The conceptual hybrid prototype proposed in this research does not operate on real user data and was developed solely for exploratory and illustrative purposes. Consequently, it raises no immediate privacy or data security concerns. Any future implementation of this system will incorporate appropriate data protection measures, transparent user consent mechanisms, and approval from relevant institutional ethical review bodies in accordance with applicable research regulations.

\section*{Acknowledgments}
We sincerely thank everyone who participated in the survey and interviews for their time and experience sharing. Also, thanks to OpenAI for providing the free version of ChatGPT, which was used to improve grammatical accuracy and language proficiency.
\section*{Funding}
This research was supported by the Institution for Advanced Research Publication Grant of United International University (Reference No.: IAR-2026-Pub-000).

\section*{Competing Interests}
The authors declare that they have no competing interests.

\makeatletter
\renewcommand{\@biblabel}[1]{#1.\hfill}
\makeatother
\appendix
\section*{Appendix}
All survey items were administered in English and were reviewed for clarity through pilot testing prior to distribution.

\section{Survey Instrument}
\label{sec:appendix-survey}
\begin{tcolorbox}[
  title=Survey Questionnaire (Selected Items),
  colback=blue!5,
  colframe=blue!70!black,
  fonttitle=\bfseries,
  boxrule=1pt,
  arc=3mm
]

\textbf{Tool Usage Frequency (Likert scale: Never to Always)}
\begin{itemize}
  \item How often do you use LLMs (e.g., ChatGPT) for English test preparation?
  \item How often do you use Google or other search engines for English test preparation?
\end{itemize}

\textbf{Perceived Satisfaction, Efficiency, and Ease of Use (Likert scale)}
\begin{itemize}
  \item How satisfied are you with the accuracy of LLMs in solving English test problems?
  \item How satisfied are you with the accuracy of search engines in solving English test problems?
  \item How efficient are LLMs in helping you complete English test tasks?
  \item How efficient are search engines in helping you complete English test tasks?
  \item How easy is it to use LLMs for English test problem solving?
  \item How easy is it to use search engines for English test problem solving?
\end{itemize}

\textbf{Tool Preference}
\begin{itemize}
  \item Which tool do you prefer overall for English test preparation?
\end{itemize}

\textbf{Open-Ended Questions}
\begin{itemize}
  \item In which English test tasks (e.g., reading, writing, grammar, vocabulary) do you prefer using LLMs over search engines, and why?
  \item What limitations or challenges have you faced while using LLMs or search engines for English test preparation?
\end{itemize}

\end{tcolorbox}

\section{Interview Tasks}
\label{sec:appendix-interview}

\begin{tcolorbox}[
  title=Assigned English Language Test-Based Tasks,
  colback=green!5,
  colframe=green!60!black,
  fonttitle=\bfseries,
  boxrule=1pt,
  arc=3mm
]

Participants were asked to complete the following English language test-related tasks using GPT-based tools, traditional search engines, or both:

\begin{enumerate}
  \item \textbf{Summarizing an english languages test based passage.} 
  Participants were asked to read  english languages test based passage and produce a concise summary that preserved the main arguments and key supporting points while maintaining perfect tone.

  \item \textbf{Correcting grammatical errors in a short text.}
  Participants reviewed  languages testing passage containing deliberate grammatical and syntactic errors and were instructed to identify and correct these errors in accordance with standard English conventions.

  \item \textbf{Selecting appropriate vocabulary in context.}
  Participants selected contextually appropriate vocabulary items to complete sentences, focusing on lexical accuracy, collocation, and semantic appropriateness commonly assessed in English language proficiency tests.

  \item \textbf{Paraphrasing sentences without altering meaning.}
  Participants paraphrased given sentences while preserving the original meaning,
  emphasizing lexical variation, syntactic restructuring, and avoidance of plagiarism,
  as required in academic writing and standardized tests.

  \item \textbf{Drafting an opinion-based response similar to IELTS Writing Task~2.}
  Participants composed a argumentative response to a prompt resembling
  IELTS Writing Task~2, focusing on coherence, clarity of argument, lexical range,
  and grammatical accuracy.

  \item \textbf{Drafting a response similar to IELTS Writing Task~1.}
  Participants wrote a brief descriptive response based on visual or textual information, mirroring IELTS Writing Task~1 requirements, with emphasis on accurate data interpretation,  organization, and appropriate use of academic language.

  \item \textbf{Interpreting information from an academic reading passage.}
  Participants answered comprehension and interpretation questions based on an academic
  reading passage, requiring identification of main ideas, inference, and understanding
  of supporting details.
\end{enumerate}

\end{tcolorbox}

\section{Task Evaluation Rubric}
\label{sec:appendix-rubric}

\begin{tcolorbox}[
  title=Rubric for Evaluating Task Performance,
  colback=orange!6,
  colframe=orange!70!black,
  fonttitle=\bfseries,
  boxrule=1pt,
  arc=3mm
]

Each task was evaluated using a predefined rubric to ensure consistency and objectivity:

\begin{itemize}
  \item \textbf{Accuracy}: Correctness of information and task completion.
  \item \textbf{Clarity}: Logical structure, coherence, and readability.
  \item \textbf{Completeness}: Coverage of required task elements.
  \item \textbf{Efficiency}: Time taken to complete the task.
  \item \textbf{Appropriateness}: Suitability of the response for an academic English test context.
\end{itemize}

Each task was independently scored by two evaluators, with disagreements resolved through discussion.

\end{tcolorbox}
\section{ Thematic Codebook}
Thematic analysis of open-ended survey responses
and interview transcripts resulted in four overarching themes. Each theme is described in Table \ref{tab:thematic-mapping}, along with its associated codes and representative
participant quotes.

\begin{table*}[t]
\centering
\resizebox{\textwidth}{!}{%
\begin{tabular}{p{3.0cm} p{3.2cm} p{5.6cm} p{5.6cm}}
\hline
\textbf{Parent Theme} & \textbf{Child Code} & \textbf{Definition} & \textbf{Example Quote} \\
\hline
Effectiveness & Task Fit &
Perceived suitability of a tool for specific English language test tasks such as writing, grammar, reading, or vocabulary practice. &
``GPT is very helpful for writing and summaries, but I still need Google when I want exact rules or examples.'' \\
\hline
Usability \& Cognitive Load & Ease of Use &
Degree to which the tool is perceived as intuitive and easy to interact with during English test preparation. &
``ChatGPT feels like a tutor and saves me time because I don’t need to open many websites.'' \\
\cline{2-4}
 & Information Overload &
User frustration caused by excessive or irrelevant information when using traditional search engines. &
``Google shows too many links, and it takes time to figure out which one is actually useful.'' \\
\hline
Trust and Credibility & Source Verification &
Need to verify information using reliable and citable sources, especially for grammar rules and writing guidance. &
``I trust Google more when I need to check if a grammar rule is actually correct.'' \\
\hline
Contextual Tool Selection & Task Type Influence &
Tool choice depends on the type and complexity of the English test task. &
``For grammar and paraphrasing I use GPT, but for reading practice and explanations I use Google.'' \\
\hline
\end{tabular}%
}
\caption{Thematic mapping of child codes derived from open-ended survey responses and in-person interviews. Example quotes illustrate typical user perceptions.}
\label{tab:thematic-mapping}

\end{table*}

\end{document}